\documentclass[aps,showpacs,showkeys]{revtex4}

\begin{document}


\newcommand\beq{\begin{equation}}
\newcommand\eeq{\end{equation}}
\newcommand\beqa{\begin{eqnarray}}
\newcommand\eeqa{\end{eqnarray}}
\newcommand\ket[1]{|#1\rangle}
\newcommand\bra[1]{\langle#1|}
\newcommand\scalar[2]{\langle#1|#2\rangle}

\newcommand\jo[3]{\textbf{#1}, #3 (#2)}


\title{\Large\textbf{Insecurity of Quantum Bit Commitment
                     with Secret Parameters}}

\author{Chi-Yee Cheung}

\email{cheung@phys.sinica.edu.tw}

\affiliation{Institute of Physics, Academia Sinica\\
             Taipei, Taiwan 11529, Republic of China\\}


\begin{abstract}

The impossibility proof of unconditionally secure quantum
bit commitment is crucially dependent on the assertion that
Bob is not allowed to generate probability distributions
unknown to Alice. This assertion is actually not
meaningful, because Bob can always cheat without being
detected. In this paper we prove that, for any concealing
protocol involving secret probability distributions, there
exists a cheating unitary transformation that is known to
Alice. Our result closes a gap in the original
impossibility proof.

\end{abstract}



\maketitle

\section{Introduction}

Bit commitment is an important primitive that can be used
to implement other two-party cryptographic protocols
\cite{Brassard-96}. In a bit commitment protocol, Alice
commits to Bob a secret bit $b\in\{0,1\}$ that is to be
unveiled at a later time. In order to guarantee that she
will not change her mind, Alice sends Bob a piece of
evidence that can later on be used to verify her honesty
when she unveils.

A bit commitment scheme is secure if (1) Bob cannot extract
the value of $b$ before Alice unveils it (concealing), and
(2) Alice cannot change the value of $b$ without Bob's
knowledge (binding). Furthermore, if the scheme remains
secure even if Alice and Bob were endowed with capabilities
limited only by the laws of nature, then it is said to be
unconditionally secure.

In a typical classical bit commitment scheme, Alice writes
the committed bit $b$ on a piece of paper and locks it in a
strong safe. She then delivers the safe to Bob but keeps
the key. Later she unveils by disclosing the bit value and
presenting the key to Bob for verification. However such a
scheme is clearly not unconditionally secure because its
security depends on, among other things, the assumption
that Bob cannot open the safe without the help of Alice. In
fact all classical bit commitment schemes are based on some
unproven assumptions, so that unconditional security is not
possible in classical settings.

By introducing quantum mechanics into the bit commitment
game, one hopes to achieve unconditional security which is
guaranteed by the laws of nature. In a quantum bit
commitment (QBC) protocol, Alice and Bob execute a series
of quantum and classical operations, such that at the end
of the commitment phase, Bob has in his hand a quantum
state characterized by a density matrix $\rho^{(b)}_{B}$.
The idea is that, with additional information from Alice in
the unveiling phase, Bob can use $\rho^{(b)}_{B}$ to check
whether Alice is honest.

\section{No-Go Theorem}

It is generally believed that Lo and Chau
\cite{Lo-97,Lo-98} and Mayers \cite{Mayers97,Brassard-97}
proved in 1997 that unconditionally secure QBC is
impossible. The arguments can be summarized as follows.
First of all, it is observed that the whole commitment
process, which may involves any number of rounds of quantum
and classical exchanges between Alice and Bob, can always
be represented by an unitary transformation ${\cal
U}^{(b)}_{AB}$ on an initial pure state
$\ket{\phi^{(b)}_{AB}}$ in the combined Hilbert space
$H_A\otimes H_B$ of Alice and Bob. Therefore at the
conclusion of the commitment process, the overall state is
given by
 \beq
 \ket{\Psi^{(b)}_{AB}}=\mathcal{U}^{(b)}_{AB}
 \,\ket{\phi^{(b)}_{AB}}.
 \eeq
The pure state $\ket{\Psi^{(b)}_{AB}}$ is called a
purification of the density matrix $\rho^{(b)}_B$ such that
 \beq
 {\rm Tr}_A~ \ket{\Psi^{(b)}_{AB}}
 \bra{\Psi^{(b)}_{AB}}
 =\rho_B^{(b)}.
 \eeq
In this approach, Alice and Bob can leave all undisclosed
parameters undetermined at the quantum level. Moreover
since the reduced density matrix $\rho^{(b)}_B$ on Bob's
side is unchanged, he cannot distinguish whether Alice
purifies or not.

In order that the protocol is concealing, the density
matrices $\rho^{(0)}_B$ and $\rho^{(1)}_B$ must be either
equal,
 \beq
 \rho^{(0)}_B = \rho^{(1)}_B,\label{perfect}
 \eeq
or arbitrarily close to each other,
 \beq
 \rho^{(0)}_B \approx \rho^{(1)}_B,\label{near-perfect}
 \eeq
corresponding respectively to the perfect concealing and
near-perfect concealing cases. The closeness between the
two density matrices, $\rho^{(1)}_B$ and $\rho^{(0)}_B$,
can be described quantitatively by the fidelity
$F(\rho_B^{(1)},\rho_B^{(0)})$.  Let
$\ket{\Phi^{(b)}_{AB}}$ be any purification of
$\rho_B^{(b)}$ so that
 \beq
 {\rm Tr}_A~ \ket{\Phi^{(b)}_{AB}}
 \bra{\Phi^{(b)}_{AB}}
 =\rho_B^{(b)}.
 \eeq
Then, according to Uhlmann's Theorem, the fidelity can be
expressed as
 \beq
 F(\rho_B^{(1)},\rho_B^{(0)})
 =\textrm{max}\,|\scalar{\Phi_{AB}^{(1)}}
 {\Phi_{AB}^{(0)}}|,
 \eeq
where the maximization is over all possible purifications,
and  $0\le F(\rho_B^{(1)},\rho_B^{(0)})\le 1$. Note that
 \beq
 F(\rho_B^{(1)},\rho_B^{(0)}) = 1
 \eeq
if and only if the perfect concealing condition, Eq.
(\ref{perfect}), holds; in this case Bob can extract
absolutely no information about Alice's committed bit $b$
from $\rho_B^{(b)}$. In general we have
 \beq
 F(\rho_B^{(1)},\rho_B^{(0)})=1-\delta,
 \eeq
where $\delta\ge 0$. For the near-perfect concealing case,
Eq. (\ref{near-perfect}), we have $\delta>0$ and it can be
made arbitrarily small by increasing the security parameter
$N$.

It is well known \cite{Nielsen-00} that for a fixed
purification $\ket{\Psi^{(1)}_{AB}}$ of $\rho^{(1)}_B$,
there exists a purification $\ket{\Phi_{AB}^{(0)}}$ of
$\rho^{(0)}_B$, such that
 \beq
 |\scalar{\Psi^{(1)}_{AB}}
 {\Phi^{(0)}_{AB}}|=
 1-\delta.
 \label{scalar}
 \eeq
Furthermore since both $\ket{\Phi^{(0)}_{AB}}$ and
$\ket{\Psi_{AB}^{(0)}}$ are purifications of the same
reduced density matrix $\rho^{(0)}_B$, they are related by
an unitary transformation:
 \beq
 \ket{\Phi_{AB}^{(0)}}=U_A \ket{\Psi^{(0)}_{AB}},
 \label{UA}
 \eeq
where $U_A$ acts on Alice's Hilbert space $H_A$ only
\cite{Nielsen-00}. In particular, for the perfect
concealing case where $\delta=0$, it is clear from Eqs.
(\ref{scalar}, \ref{UA}) that
 \beq
 U_A\ket{\Psi_{AB}^{(0)}}=\ket{\Psi^{(1)}_{AB}},
 \eeq
apart from an unimportant phase factor.

The existence of $U_A$ means that Alice can cheat with the
following strategy (called EPR attack). To begin with, she
always commits to $b=0$. Later on, right before she
unveils, if she wants to keep her initial commitment, she
simply follows the protocol honestly to the end. Otherwise
if she wants to switch to $b=1$, she only needs to apply
$U_A$ to her share of the state $\ket{\Psi^{(0)}_{AB}}$,
and then proceed as if she had committed to $b=1$ in the
first place. In the perfect concealing case, Alice succeeds
with probability one. Otherwise, in the near-perfect case,
her success probability approaches unity as
$N\rightarrow\infty$ ($\delta\rightarrow 0)$. Hence if a
protocol is concealing, it cannot be binding at the same
time.  This is the no-go theorem of unconditionally secure
quantum bit commitment
\cite{Lo-97,Lo-98,Mayers97,Brassard-97}.

\section{Secret Parameters}

It has been pointed out that the above proof only
establishes the existence of the cheating transformation
$U_A$, but there is no guarantee that $U_A$ is always known
to Alice \cite{Cheung05,Yuen03}. The point is, even in the
fully purified approach, the overall state
$\ket{\Psi^{(b)}_{AB}(\omega)}$ may still depend on some
probability distribution $\omega$ unknown to Alice. If so,
then the cheating transformation $U_A(\omega)$ would in
general depend on $\omega$, and Alice would not be able to
implement $U_A(\omega)$ without the help of Bob. This is a
serious logical gap in the original impossibility proof. To
overcome this gap, the proof
\cite{Lo-97,Lo-98,Mayers97,Brassard-97} asserts that Alice
knows in detail all the probability distributions generated
by Bob in any QBC protocol, hence she knows $U_A(\omega)$.

This assertion is actually not correct. As shown in the
Appendix, it is not meaningful to specify a probability
distribution to an untrustful party (Bob) in a quantum
protocol, because he can always cheat without being
detected \footnote{See also Appendix of Ref.
\cite{Cheung05}, v2.}. So, regardless of whether secret
parameters are allowed in QBC protocols or not, they are
potentially there and must be taken into account in
security analysis. Consequently, whether the no-go theorem
remains valid in the presence of secret parameters is a
crucial question that cannot be avoided and has yet to be
answered.

In Ref. \cite{Cheung05} it is shown that, in the perfect
concealing case ($\rho^{(0)}_B=\rho^{(1)}_B$), Alice can
cheat and succeed for sure without knowing Bob's secret
choices. In this paper, we present a general proof that
unconditionally secure QBC is impossible even if Bob is
allowed to generate probabilities unknown to Alice.
Specifically we shall prove that, for any perfect or
near-perfect concealing QBC protocol involving a secret
probability distribution $\omega$ unknown to Alice, there
exists a cheating unitary transformation independent of
$\omega$ with which Alice can cheat.

Consider first the near-perfect case. Suppose we are given
a protocol which is proven to be near-perfect concealing
for whatever secret $\omega$ Bob chooses to use. Let
 \beq
 \omega=\{q_1,\ldots,q_m\},
 \label{omega}
 \eeq
where $q_j\ge 0$ and
 \beq
 \sum^m_{j=1} q_j=1;
 \eeq
otherwise the $q_j$'s are arbitrary and unknown to Alice.
Let $\Omega^*$ be a special set of distributions:
 \beq
 \Omega^*=\{\omega^*_1,\ldots,\omega^*_m\},
 \label{Omega*}
 \eeq
where
 \beq
 \omega^*_j=\{0,\ldots,q_j=1,\ldots,0\}.
 \eeq
The near-perfect concealing property implies that
 \beq
 F(\rho^{(1)}_B(\omega^*_j),
 \rho^{(0)}_B(\omega^*_j))=1-\delta^*_j,
 \eeq
where $\delta^*_j>0$, and $\delta^*_j\rightarrow 0$
asymptotically as the security parameter $N\rightarrow 0$
for all $\omega^*_j$ in $\Omega^*$. It then follows from
previous arguments that, for each $\omega^*_j$, there
exists a cheating unitary transformation $U_A(\omega^*_j)$,
such that
 \beq
 |\bra{\Psi^{(1)}_{AB}(\omega^*_j)}
 U_A(\omega^*_j) \ket{\Psi^{(0)}_{AB}(\omega^*_j)}|
 =1-\delta^*_j, \label{delta*}
 \eeq
where $U_A(\omega^*_j)$ depends on $\omega^*_j$ in general.

Since $\omega$ is not revealed to Alice, Bob can purify his
options with an arbitrary probability distribution over any
set of possible choices. Consider the following
purification over $\Omega^*$,
 \beq
 \ket{\Psi'^{(b)}_{AB}}=\sqrt{1/m}\,\sum_{j=1}^m\,
 \ket{\Psi^{(b)}_{AB}(\omega^*_j)}\ket{\xi_j},
 \label{Psi'}
 \eeq
where $\{\ket{\xi_j}\}$ is a set of orthonormal ancilla
states.
The corresponding reduced density matrix,
 \beq
 \rho'^{(b)}_B=\textrm{Tr}_A
 \ket{\Psi'^{(b)}_{AB}}\bra{\Psi'^{(b)}_{AB}},
 \eeq
should also satisfy the near-perfect concealing condition
 \beq
 F(\rho'^{(1)}_B, \rho'^{(0)}_B)=1-\delta',
 \eeq
where $\delta'>0$, and $\delta'\rightarrow 0$ as
$N\rightarrow\infty$. Hence, as explained before, there
exists a cheating unitary transformation $U'_A$, such that
 \beq
 \bra{\Psi'^{(1)}_{AB}} U'_A
 \ket{\Psi'^{(0)}_{AB}}= 1-\delta',
 \label{delta'}
 \eeq
where the phase factor has been absorbed into $U'_A$ for
convenience. Notice that $U'_A$ is independent of any
secret parameters, so it is known to Alice.  We shall show
that Alice can use this $U'_A$ to cheat, no matter how Bob
purifies his secret choice of $\omega$.

Substituting Eq. (\ref{Psi'}) into Eq. (\ref{delta'}), we
get
 \beq
 \frac{1}{m}\sum_{j=1}^m
 \bra{\Psi^{(1)}_{AB}(\omega^*_j)}U'_A
 \ket{\Psi^{(0)}_{AB}(\omega^*_j)}
 =1-\delta'.
 \label{UA'}
 \eeq
Let
 \beq
 \bra{\Psi^{(1)}_{AB}(\omega^*_j)}U'_A
 \ket{\Psi^{(0)}_{AB}(\omega^*_j)}=(1-\alpha_j)+i\beta_j,
 \label{alpha-beta}
 \eeq
where $\alpha_j$ and $\beta_j$ are real, and $\alpha_j>0$;
then one can show that $\delta'\rightarrow 0$ if and only
if every $\alpha_j\rightarrow 0$ and $\beta_j\rightarrow
0$. Intuitively this must be true because the two vectors,
$\ket{\Psi'^{(1)}_{AB}}$ and $U'_A\ket{\Psi'^{(0)}_{AB}}$,
can be nearly identical if and only if the corresponding
orthogonal components, $\ket{\Psi^{(1)}_{AB}}(\omega^*_j)$
and $U'_A\ket{\Psi^{(0)}_{AB}}(\omega^*_j)$, are all nearly
identical. This statement can be made quantitative as
follows. Substituting Eqs. (\ref{alpha-beta}) into Eq.
(\ref{UA'}), we get
 \beq
 \delta'=\frac{1}{m}\,\sum_{j=1}^m \alpha_j,
 \label{sum-alpha}
 \eeq
and
 \beq
 \sum_{j=1}^m\beta_j=0.
 \eeq
Eq. (\ref{sum-alpha}) shows that $\delta'\rightarrow 0$ if
and only if all $\alpha_j\rightarrow 0$ as
$N\rightarrow\infty$; furthermore each $\alpha_j$ should
approach zero at least as fast as $\delta'$. Hence
$\alpha_j$ must satisfy
 \beq
 \alpha_j \le c\delta',\label{c1}
 \eeq
where $0<c\le m$ is a constant independent of $N$. The fact
that
 \beq
 |\bra{\Psi^{(1)}_{AB}(\omega^*_j)}U'_A
 \ket{\Psi^{(0)}_{AB}(\omega^*_j)}| < 1
 \eeq
implies
 \beq
 (\alpha_j^2+\beta_j^2)/2 < \alpha_j \le c\delta',
 \label{c2}
 \eeq
hence $\beta_j\rightarrow 0$ as $\alpha_j\rightarrow 0$.
Then we have
 \beqa
 |\bra{\Psi^{(1)}_{AB}(\omega^*_j)}U'_A
 \ket{\Psi^{(0)}_{AB}(\omega^*_j)}|^2
 &=& 1-2\alpha_j+\alpha_j^2+\beta_j^2,
 \nonumber\\
 &>& 1-2c\delta',
 \eeqa
This result shows that, for any $\omega^*_j$ in $\Omega^*$,
Alice can use $U'_A$ to cheat and her success probability
is arbitrarily close to unity. That means, for practical
purpose, Alice can use $U'_A$ in place of the optimal but
unknown $U_A(\omega^*_j)$ in Eq. (\ref{delta*}), even
though the two transformations may not be exactly equal.

Next we show that Alice can use $U'_A$ to cheat even if Bob
uses an arbitrary $\omega$ as given in Eq. (\ref{omega}).
By definition, $\ket{\Psi^{(b)}_{AB}(\omega)}$ is a
purification over the set $\Omega^*$ [see Eq.
(\ref{Omega*})], \textit{viz.},
 \beq
 \ket{\Psi^{(b)}_{AB}(\omega)}=
 \sum_{j=1}^m\sqrt{q_j}\,
 \ket{\Psi^{(b)}_{AB}(\omega^*_j)}\ket{\xi_j}.
 \label{Psi-omega}
 \eeq
Therefore according to Eq. (\ref{alpha-beta}),
 \beqa
 \bra{\Psi^{(1)}_{AB}(\omega)}U'_A
 \ket{\Psi^{(0)}_{AB}(\omega)}
 &=&\sum_{j=1}^m q_j
 \bra{\Psi^{(1)}_{AB}(\omega^*_j)}U'_A
 \ket{\Psi^{(0)}_{AB}(\omega^*_j)},\nonumber\\
 &=& 1-\bar\alpha + i\bar\beta,
 \label{alpha-beta-bar}
 \eeqa
where
 \beqa
 \bar\alpha&=&\sum_{j=1}^m q_j\alpha_j,\label{alpha-bar}\\
 \bar\beta&=&\sum_{j=1}^m q_j\beta_j.\label{beta-bar}
 \eeqa
From Eq. (\ref{c1}) and Eq. (\ref{alpha-bar}), we get
 \beq
 \bar\alpha \le c\delta',
 \eeq
which, together with
 \beq
 |\bra{\Psi^{(1)}_{AB}(\omega)}U'_A
 \ket{\Psi^{(0)}_{AB}(\omega)}| < 1,
 \eeq
gives
 \beq
 (\bar\alpha^2+\bar\beta^2)/2 < \bar\alpha \le c\delta'.
 \label{alpha-bar^2}
 \eeq
Then
 \beqa
 |\bra{\Psi^{(1)}_{AB}(\omega)}U'_A
 \ket{\Psi^{(0)}_{AB}(\omega)}|^2
 &=& 1-2\bar\alpha+\bar\alpha^2+\bar\beta^2,\nonumber\\
 &>&1-2c\delta' \label{probability}.
 \eeqa
Consequently Alice can use $U'_A$ to cheat, independent of
what $\omega$ Bob chooses to use. We emphasize that $U'_A$
may not necessarily maximize the quantity
$|\bra{\Psi^{(1)}_{AB}(\omega)}U'_A
\ket{\Psi^{(0)}_{AB}(\omega)}|$, nevertheless Eq.
(\ref{probability}) shows that Bob can use it to achieve
the cheating purpose for arbitrary $\omega$.

Finally we show that this same $U'_A$ also works if Bob
purifies his choices over an arbitrary set of $\omega$'s,
$\Omega=\{\omega_1,\ldots,\omega_n\}$, where
 \beq
 \omega_k=\{q^k_1,\ldots,q^k_m\}
 \eeq
as shown in Eq. (\ref{omega}). A general purification over
$\Omega$ can be written as
 \beq
 \ket{\Psi''^{(b)}_{AB}}=\sum_{k=1}^n\,\sqrt{p^{}_k}\,
 \ket{\Psi^{(b)}_{AB}(\omega_k)}\ket{\chi^{}_k},
 \eeq
where $\ket{\Psi^{(b)}_{AB}(\omega_k)}$ is given by Eq.
(\ref{Psi-omega}), $\ket{\chi^{}_k}$'s are orthonormal
ancilla states, and $\{p^{}_1,\ldots,p^{}_n\}$ is any
probability distribution such that
 \beq
 \sum_{k=1}^n p^{}_k=1.
 \eeq
Then following the arguments presented earlier, we get
 \beq
 |\bra{\Psi''^{(1)}_{AB}}U'_A
 \ket{\Psi''^{(0)}_{AB}}|^2
 > 1-2c \delta'.
 \label{probability-2}
 \eeq
This result can also be easily obtained as follows. By a
redefinition of the ancilla states, we can rewrite
$\ket{\Psi''^{(b)}_{AB}}$ in terms of a single effective
distribution \cite{Cheung05}:
 \beq
 \ket{\Psi''^{(b)}_{AB}}=
 \ket{\Psi^{(b)}_{AB}(\omega'')},
 \eeq
where $\omega''=\{q''_1,\ldots,q''_m\}$ is given by
 \beq
 q''_j=\sum^n_{k=1} p^{}_k q^k_j.
 \eeq
Then Eq. (\ref{probability-2}) follows directly from Eq.
(\ref{probability}). Thus we conclude that, for any
near-perfect concealing QBC protocol, Alice can use $U'_A$
of Eq. (\ref{delta'}) as the cheating transformation, no
matter how Bob purifies his secret choices. In all cases,
she succeeds with a probability $P_A(N)$ that can be made
arbitrarily close to one by increasing the security
parameter $N$.

It is straightforward to extend the above proof to cover
the perfect concealing case as well. The perfect concealing
condition, Eq. (\ref{perfect}), implies that
 \beq
 \delta^*_j=0
 \eeq
in Eq. (\ref{delta*}), and
 \beq
 \delta'=0
 \eeq
in Eq. (\ref{delta'}). It then follows from Eq. (\ref{UA'})
that
 \beq
 \bra{\Psi^{(1)}_{AB}(\omega^*_j)}U'_A
 \ket{\Psi^{(0)}_{AB}(\omega^*_j)} = 1
 \eeq
for all $\omega^*_j\in\Omega^*$. Hence
 \beq
 \alpha_j=\beta_j=0
 \eeq
in Eq. (\ref{alpha-beta}), and
 \beq
 \bar\alpha=\bar\beta=0
 \eeq
in Eq. (\ref{alpha-beta-bar}). The above results imply that
 \beq
 U'_A=U_A(\omega)
 \eeq
for arbitrary $\omega$, and the success probability
$P_A(N)=1$. Therefore if $\rho_B^{(0)}=\rho_B^{(1)}$, then
Alice can use $U'_A$ to cheat and succeed with probability
equal to one, independent of Bob's secret choices.

Finally we note that the question of whether $U'_A$ depends
on Bob's ancilla states has also been raised \cite{Yuen03}.
The fact that it does not can be seen as follows. We know
that any two different sets of ancilla states on Bob side
are related by an unitary transformation $U_B$ acting on
Bob's Hilbert space $H_B$. Since
 \beq
 [U_B,U'_A]=0,
 \eeq
it is obvious that $U'_A$ does not depend on the particular
ancilla set Bob chooses to use.

\section{Conclusion}

In this paper we have proved that, for any perfect or
near-perfect concealing QBC protocol involving a
probability distribution $\omega$ unknown to Alice, there
exists an $\omega$-independent unitary transformation with
which Alice can cheat. Our result closes a gap in the
original impossibility proof
\cite{Lo-97,Lo-98,Mayers97,Brassard-97}. We conclude that,
for those protocols covered by the original proof,
unconditionally secure QBC is impossible even if Bob
employs secret parameters.

\vspace{3em}

\centerline{\textbf{APPENDIX}}

\vspace{1em}

Suppose a protocol specifies that Bob should take certain
action $V_j\,(j=1,\ldots,m)$ on a state $\ket{\phi}$,
according to a probability distribution
$\omega_0=\{q^0_1,\ldots,q^0_m\}$. In the purified form,
the resultant state is given by
 \beq
 \ket{\psi(\omega_0)}=\sum_{j=1}^m \sqrt{q^0_j}\,
 \ket{\xi_j}\,V_j\,\ket{\phi},
 \eeq
where $\ket{\xi_j}$'s are orthonormal ancilla states. As
shown in Ref. \cite{Cheung05}, a superposition of
$\ket{\psi(\omega_k)}$'s, where
$\omega_k=\{q^k_1,\ldots,q^k_m\}$, can effectively be
written in terms of a single distribution, \textit{i.e.},
 \beqa
 \ket{\psi'}&=&\sum_{k=1}^n\sqrt{p^{}_k}\,
 \,\ket{\chi^{}_k}\ket{\psi(\omega_k)} \label{entangle2}\\
 &=& \ket{\psi(\omega')},
 \eeqa
where $\ket{\chi^{}_k}$'s are ancilla states,
$\{p_1,\ldots,p_n\}$ is a probability distribution, and
$\omega'=\{q'_1,\ldots,q'_m\}$ is the effective
distribution given by
\beq
 q'_j=\sum^n_{k=1} p^{}_k q_j^k.
 \eeq
Let $\omega'=\omega_0$, then it is clear that Bob could
generate $\ket{\psi'}$ instead of $\ket{\psi(\omega_0)}$,
and he would have no problem passing any possible checks
initiated by Alice. In general some qubits are measured and
discarded in the checking procedure. For the remaining
qubits, Bob could either stay with $\omega_0$, or he could
collapse the ancillas $\{\ket{\chi_k}\}$ in Eq.
(\ref{entangle2}) to obtain a state $\ket{\psi(\omega_i)}$,
where $\omega_i$ is not equal to $\omega_0$ in general.

Hence it is not meaningful for Alice to specify a
probability distribution to an untrustful Bob, because
there is no way to enforce it.


\acknowledgments  The author thanks H. P. Yuen for useful
discussions and comments.



\end{document}